\documentclass[journal=jacsat,manuscript=article]{achemso}

\usepackage{xcolor}

\author{Gonca Erdemci-Tandogan}
 \affiliation{Department of Physics and Astronomy,
   University of California, Riverside, California 92521, USA}
 \author{Jef Wagner}
 \affiliation{Department of Physics and Astronomy,
   University of California, Riverside, California 92521, USA}
 \author{Paul van der Schoot}
 \affiliation{Group Theory of Polymers and Soft Matter, Eindhoven University of Technology, P.O. Box 513, 5600 MB Eindhoven,
   The Netherlands}
 \alsoaffiliation{ Institute for Theoretical Physics,
   Utrecht University,
   Leuvenlaan 4, 3584 CE Utrecht, The Netherlands}
\author{Roya Zandi}
 \affiliation{Department of Physics and Astronomy,
   University of California, Riverside, California 92521, USA}
\email{roya.zandi@ucr.edu}

\title[An \textsf{achemso} demo]{Role of Genome in the Formation of Conical Retroviral Shells}

\begin{document}

\begin{abstract}
Human immunodeficiency virus (HIV) capsid proteins spontaneously assemble around the genome into
a protective protein shell called the capsid, which can take on
a variety of shapes broadly classified as conical, cylindrical and
irregular. The majority of capsids seen in {\it in vivo} studies are conical in
shape, while {\it in vitro} experiments have shown a
preference for cylindrical capsids. The factors involved in the selection of the unique shape of
HIV capsids are not well understood, and in particular the impact of RNA on the formation of the capsid is not known. In this work, we study the role of the genome and its
interaction with the capsid protein by modeling the genomic RNA through
a mean-field theory. Our results show that the confinement free energy for
a homopolymeric model genome confined in a conical capsid is lower than that in a cylindrical capsid, at least
when the genome does not interact with the capsid, which seems to be the case in {\it in vivo} experiments. Conversely, the confinement free energy for the cylinder is lower than for a conical capsid if the genome is attracted to the capsid proteins as the {\it in vitro} experiments. Understanding the factors that contribute to the formation of conical capsids may shed light on the infectivity of HIV particles.
\end{abstract}



\maketitle

\section{INTRODUCTION}

Human immunodeficiency virus (HIV) has, not surprisingly, attracted significant attention from the wider scientific community due its association with the AIDS pandemic \cite{Adamson2010}. The lifecycle of HIV involves the budding of the non-infectious immature virion that subsequently transforms into an infectious, mature virus \cite{Ganser2008,Sundquist2012,Freed2015}. Like many other viruses, the mature HIV particle consists of a protein shell that surrounds its genetic materials, enveloped by a bilayer lipid membrane. The shell of the immature virion is roughly spherical and is built up from a large number of multi-domain protein called Gag. When the virus matures, the Gag protein is cleaved into three structural domains called matrix (MA), capsid (CA) and nucleocapsid (NC)  \cite{Ganser2008,Sundquist2012,Freed2015}. The MA protein remains attached to the membrane envelope, while the positively charged NC proteins bind to the genome. The CA protein self assembles around the genome/NC complexes to form predominantly conical capsid shells, but cylindrical and irregular shapes have also been observed\cite{Benjamin2005}. The physical processes that control the unique shapes of the mature capsid are not well understood.

Several {\it in vitro} experiments have shown that CA proteins can assemble spontaneously in solution to form cylindrical and conical shells in the absence of genome\cite{Ehrlich1992,Campbell1995,Gross1997,Gross1998,vonSchwedler1998,Ganser1999,Grattinger1999,Li2000,Pornillos2009,Byeon2009,Meng2012,Zhao2013}. {In 1999, Ganser {\it et al.} \cite{Ganser1999} used a recombinant CA-NC fusion proteins to test the impact of specific genome sequences on the formation of conical capsids by mixing the CA-NC protein not only with the native HIV genome (a 1400 nucleotide template) but also with other viral and non-viral RNAs. They found that the HIV genome is not specifically required for the formation of conical capsids.} In fact, their studies show that under high-salt concentrations both conical and cylindrical shells form, confirming that the genome is not required for the cone formation. Still, according to their experiments, the majority of the capsids that form are cylindrical, in contrast to HIV capsids assembled \textit{in vivo} \cite{Ganser1999}. We note that in the experiments of Ganser {\it et al.} RNA interacts with the shell, due to the attractive interaction between the genome and the positive charges on the recombinant CA-NC fusion proteins. On the other hand, in {\it in vivo} experiments the protease cleaves the link between NC and CA and as such the shell is solely built from CA and the positively charged NC proteins are attracted to RNA and condenses it.  It is well-known that there is no attractive interaction between RNA and CA proteins \cite{Briggs2012,Woodward2015}. 

There are a number of theoretical and computer simulation studies that investigate factors that contribute to the formation of conical capsids. The focus of prior work is on the elastic energy of the capsids, mostly in the absence of genome, and also on role of kinetic pathways towards assembly. In 2005, Nguyen, Gelbart and Bruinsma did the first equilibrium studies of HIV capsids and compared the elastic energy of conical and cylindrical shells \cite{Nguyen2005}. In their work, the presence of the genome was implicitly taken into account by putting constraints on the shells. They found that the conical capsid constitutes an elastic minimum energy structure only under the condition of fixed area, volume and spanning length. The volume constraint mimics the RNA being confined inside the capsid. The spanning length constraint is justified because {\it in vivo} the height of cylindrical and conical HIV capsids is fixed as a result of the enclosing membrane (the envelope). It is worth mentioning that cylinders and cones with the same area and volume also have the same height, that is, to lowest order in the cone aperture angle.

Despite intense research, including the aforementioned studies, the predominance of conical over cylindrical capsids remains poorly understood. In particular, no theoretical studies have explicitly taken into account the impact of the genome on the formation of retroviral capsids. In this paper, using a simple mean-field model, we obtain the free energy of confinement of a chain confined in conical and cylindrical capsids. Due to the absence of sufficient information on the CA-CA interaction and the free energies associated with it, we focus entirely on role of genome-shell interaction in the formation of capsids. We obtain the concentration profile of the model genome, a linear polymer, inside the capsid and consider the cases in which the genome interacts attractively with the capsid mirroring the experiments of Ganser {\it et al.}, and also consider {\it in vivo} conditions in which the genome presumably does not interact with the capsid wall built from CA proteins.

We find for a genome that does not interact attractively with the shell, the free energy cost of confining this genome in a conical capsid is smaller than that associated with confining it in a cylindrical one, presuming they have the same height and area mimicking the impact of the envelope and fixing the number of Gag proteins that make up the shell. In conical capsids the genome is primarily located in the base of the cone, consistent with the experimental findings of Refs. \cite{Briggs2012,Woodward2015,Steven2015,Briggs2003}. However, if the genome interacts attractively with the capsid wall, as is the case in the experiments of Ganser {\it et al.}  \cite{Ganser1999}, we find that with the same amount of genomic material and the same number of capsid proteins, the genome confinement free energy is slightly larger for a cone-shaped than that of a cylindrical capsid. This may explain why cones are less favorable than cylinders in {\it in vitro} HIV-1 experiments.

In what follows, we first present our mean-field model, which allows us to calculate the free energy of a linear genome in both cylindrical and conical capsids. We then numerically obtain the free energy of confinement of genome in the absence of genome-capsid attractive interaction but with excluded volume interaction. Next, we solve free energy of confinement in cylindrical and conical shells analytically in the absence of excluded-volume interactions.  Finally, we calculate the free energy of a genome confined in a capsid that interacts attractively with the capsid. We summarize our work and present our main conclusions in the last section.

\section{Model} \label{nointeraction}
\label{model}

Mean-field theory has been extensively used to obtain the free energy of chains confined in spherical shells\cite{Paul2005,Belyi2006,Paul2007,siber2007,Lee-Nguyen2008,Siber2008,Paul2009,Siber2010,Paul2013,Erdemci2014,Wagner2015,Klug2006,Twarock2013}. Since the focus of this paper is on HIV particles, we employ Edwards-de Gennes-Lifshitz theory to calculate the free energy of a linear polymer confined in a cylindrical and conical shell \cite{Doi1986,deGennes79,Hone1988}. In the ground-state approximation of long chains, it reads
\begin{equation}\label{FE}
\beta \Delta F = \int dV \bigg{(} \frac{a^2}{6}(\nabla \psi)^2+\frac{1}{2}\upsilon \psi^4 \bigg{)}
\end{equation}
where $a$ is the Kuhn length, $\upsilon$ is the excluded volume and $\beta=1/k_BT$ the reciprocal thermal energy with $T$ the temperature and $k_B$ the Boltzmann constant.  { The quantity $\psi$ indicates the polymer density field and $\psi^2$ represents the monomer density at positions $r$, the radial distance from the center of the capsid, and $h$, the distance along the height of the capsid. To obtain the genome profile inside the shell, we will make contour plots of $\psi^2$ as a function of $r$ and $h$.  The gradient term in Eq.~\ref{FE} is equal to zero for a constant $\psi$ and thus it is associated with the entropic cost due to the non-uniform chain distribution. } The second term is the energy penalty related to the excluded volume interaction. Hence, our reference free energy is that of a uniformly distributed polymer. In our closed system the number of encapsulated monomers, $N$, is fixed implying that

\begin{equation}\label{totalnumber}
N=\int dV \psi^2=constant.
\end{equation}

\begin{figure}
\begin{center}
\includegraphics[height=8cm]{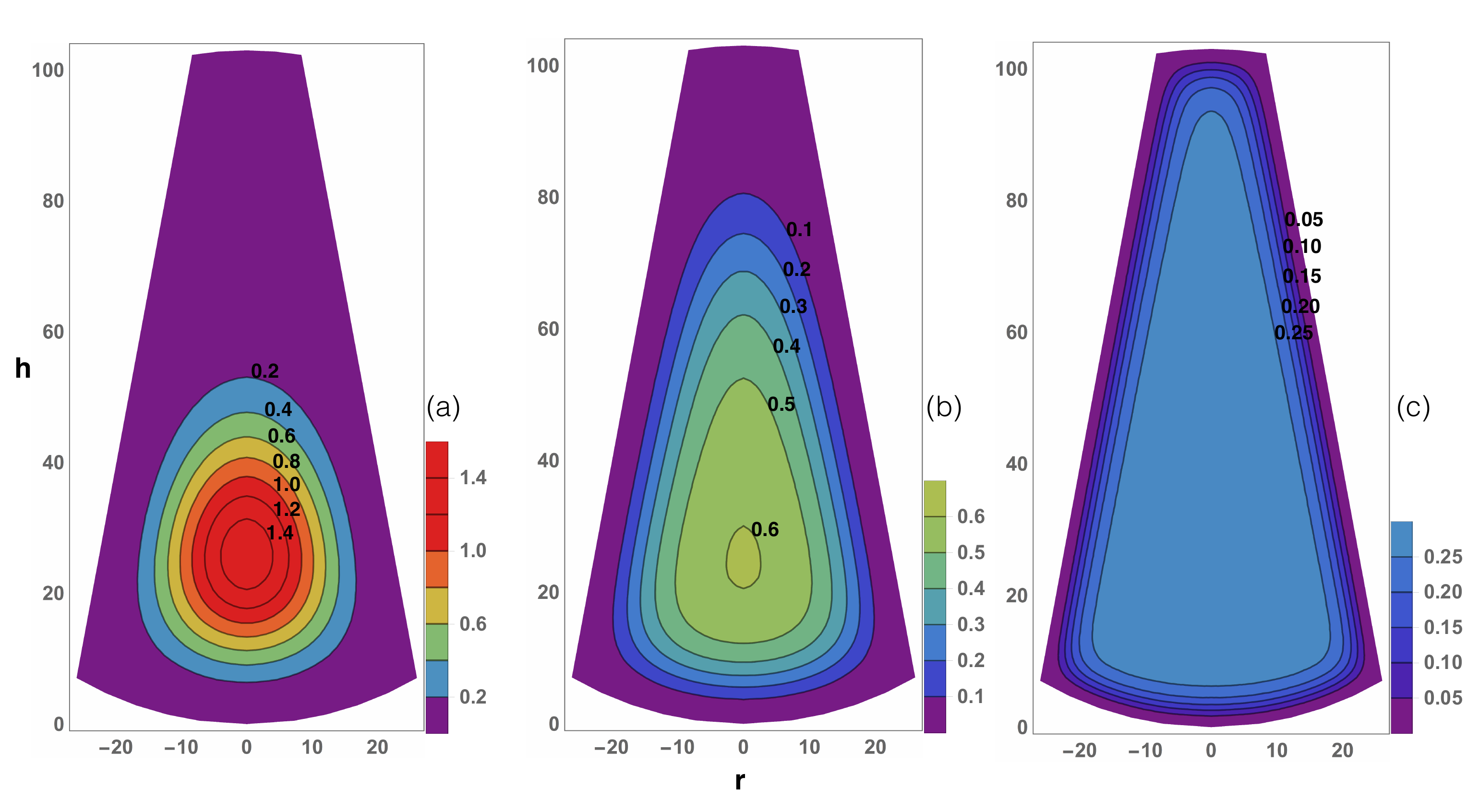}
\caption{Contour plot of the genome density profile ($\psi^2$ in units of $1/{nm^3}$) for excluded volume $\nu=0$ (a), $\nu=0.01 \;nm^3$ (b) and $\nu=0.1 \;nm^3$ (c). The total number of monomers is $N=18000$.  The other parameters are $h=103\;nm$, $R_b=26\;nm$, $\alpha=21^{\circ}$ and $a=1\;nm$. As the excluded volume increases, the polymer spreads more evenly over the cavity.}
\label{density-dirichlet-cone}
\end{center}
\end{figure}

Minimizing the free energy given in Eq.~\ref{FE} with respect to the field $\psi$, subject to the above constraint, gives the following Euler-Lagrange differential equation for the density profile inside the capsid
\begin{equation}\label{diffEq}
\frac{a^2}{6}\nabla^2\psi=-\lambda \psi+\upsilon \psi^3,
\end{equation}
where $\lambda$ is a Langrange multiplier that will be fixed by the condition of the conservation of mass.

\begin{figure}
\begin{center}
\includegraphics[height=8cm]{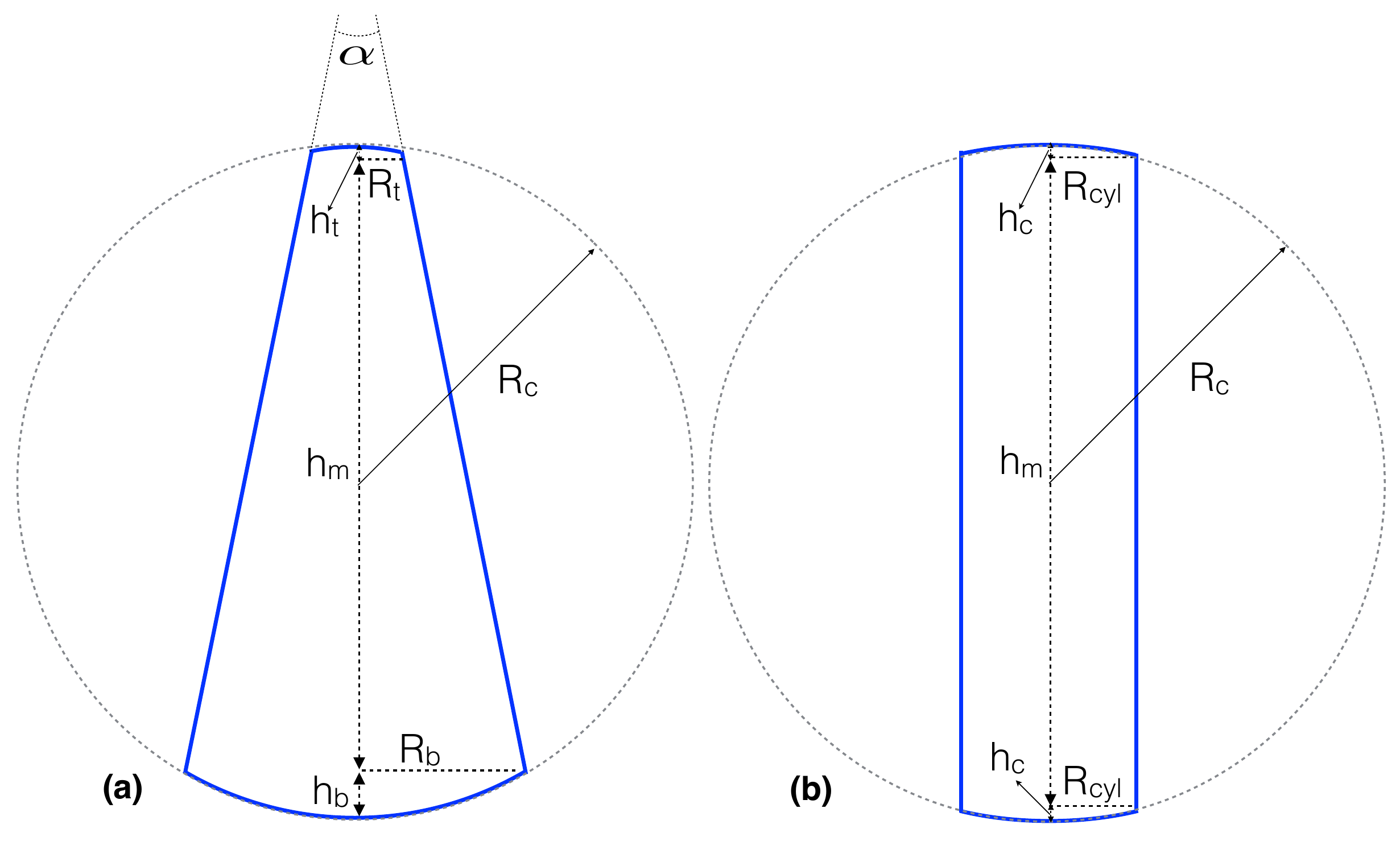}
\caption{A conical capsid (blue solid lines) inside a membrane (gray dashed sphere). It consists of the lateral surface of a truncated cone, along with two spherical caps. The spherical caps both follow the surrounding membranes, and have radii of $R_c$. The top and bottom radii of the truncated cone are given as $R_t$ and $R_b$, respectively, and the cone angle is given by $\alpha$. The perpendicular height of the cone is $h$ ($h=2 R_c$), whereas the top and bottom caps have heights of $h_t$ and $h_b$, respectively (a). The cylindrical capsid used in the calculations shares the same geometry as the conical capsid with the conditions $R_{cyl}=R_t = R_b$ and $h_t=h_b=h_c$ (b).}
\label{numerics-geometry}
\end{center}
\end{figure}

Considering there is no known interaction between the genome and capsid proteins (CA) in HIV-1 virions, we employ the Dirichlet boundary condition in which the density field at the surface $\psi \big{|}_S=0$ \cite{Doi1986}. 

{To obtain the density field $\psi$ for a polymer trapped in conical and cylindrical geometries, we numerically find the solutions of  the non-linear differential equation, Eq.~\ref{diffEq}, using a 2D Finite Element Method \cite{Gershenfeld}.  More specifically, for a given monomer number, $N$, we vary the Lagrange multiplier $\lambda$ and self-consistently solve Eqs.~\ref{totalnumber} and \ref{diffEq} till we obtain the total number of monomers equal to the one reported in the experiments and thus find the corresponding $\psi$. } Inserting the density field $\psi$ back into Eq.~\ref{FE}, we can calculate the relevant free energies as a function of the different system parameters such as the length of the polymer and the dimensions and shape of the cavity, see Fig.~\ref{numerics-geometry}. 

The results of our numerical calculations for the conical core are presented as a contour plot of the polymer segment density in Fig.~\ref{density-dirichlet-cone} with (a) a zero and (b, c) a non-zero excluded volume parameter, $\upsilon$. We set the height of the cone at $h=103 \; nm$, the radius of the base at $R_b=26 \; nm$ and presumed a cone angle of $\alpha=21^{\circ}$, consistent with observations on HIV \cite{Ganser1999,Welker2000,Briggs2003,Benjamin2005}. See also Fig.~\ref{numerics-geometry} (a).  The quantities $h_b=7.05 \; nm$ and $h_t=0.68 \; nm$  can then be easily obtained as a function of $h$, $R_b$ and $\alpha$. The Kuhn length a is $1\;nm$ in all our calculations \cite{Yoffe2008}, and the total number of monomers is chosen as $N=18000$, which is the approximate number of nucleotides in two copies of the HIV genome carried in the capsid \cite{Petropoulos1997}. The genome profile in Fig.~\ref{density-dirichlet-cone}(a) with $\nu=0$ is condensed towards the base of the cone similar to the genome profile in conical cores observed in experiments \cite{Briggs2012,Woodward2015,Steven2015,Briggs2003}. Note that due to the electrostatic interaction between the positively charged cleaved NC proteins and the negatively charged genome the solution can in some sense be considered near the $\theta$ conditions. For simplicity, thus we set the excluded volume interaction to zero, $\nu=0$. If $\nu > 0$, the chain is not as condensed at the base but is distributed more uniformly along the cone as can be seen in  Fig.~\ref{density-dirichlet-cone} (b) and (c). As illustrated in the figure, the chain locates further from the base of the cone and distributes more uniformly along the cone as $\nu$ increases.

\begin{figure}
\begin{center}
\includegraphics[height=8cm]{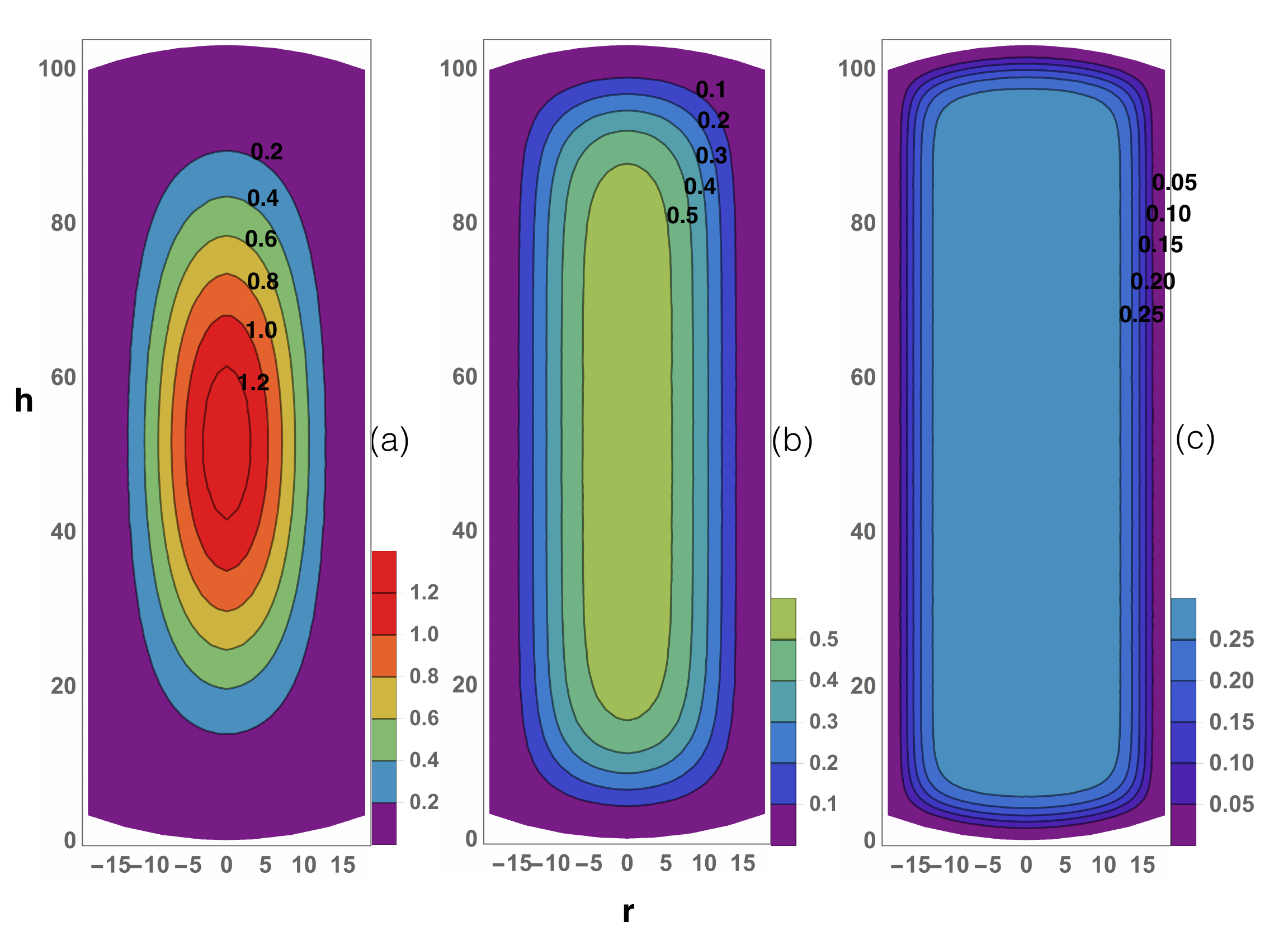}
\caption{ Contour plot of the genome density profile ($\psi^2$ in units of $1/{nm^3}$) for excluded volume $\nu=0$ (a), $\nu=0.01 \;nm^3$ (b) and $\nu=0.1 \;nm^3$ (c).  The total number of monomers is $N=18000$. Other parameters are $h=103\;nm$, $R_{cyl}=17.9\;nm$ and $a=1\;nm$. As the excluded volume increases, polymer spreads more evenly over the volume of the cavity.}
\label{density-dirichlet-cyl}
\end{center}
\end{figure}

We now compare the free energy of a genome confined in a cylindrical shell with that in a conical one as explained above (Fig.~\ref{density-dirichlet-cone}), assuming that the number of subunits are the same in both cylindrical and conical structures, and that the height of the cylinders and cones are also the same. This is reasonable because the height of the core of mature HIV-1 virions is defined by the presence of an enclosing membrane, as already advertised. Under these constraints, the radius of the cylinder becomes $R_{cyl}=17.9\;nm$ and the height of the spherical caps $h_c=3.22\;nm$, see Fig.~\ref{numerics-geometry} (b). Figure~\ref{density-dirichlet-cyl}(a) and (b, c) show the density profiles of the genome in a cylindrical core with and without monomer-monomer excluded-volume interactions, respectively.

\begin {table}
\caption {Confinement free energy for cylindrical and cone-shaped cavities. The total number of monomers is $N=18000$. Other parameters take the values $h=103\;nm$, $R_b=26\;nm$, $\alpha=21^{\circ}$, $R_{cyl}=17.9\;nm$ and $a=1\;nm$.}
\begin{center}
\begin{tabular}{ |p{2.5cm}|p{2.5cm}|p{2.5cm}| }
 \hline
 \multicolumn{3}{|c|}{$\Delta$ F ($k_B T$)} \\
 \hline
 & Conical & Cylindrical \\
 \hline
$\nu=0$ &  45  & 57  \\
\hline
$\nu=0.01 \;nm^3$ &  85  & 93  \\
 \hline
 $\nu=0.1 \;nm^3$ &  301  & 314  \\
 \hline
\end{tabular}
\label{table}
\end{center}
\end {table}

Table~\ref{table} summarizes the free energy values in units of $k_B T$ with and without excluded-volume interaction for both geometries. We find that for the same number of monomers and the same surface area, the latter imposing equal numbers of capsid proteins in the shells, the genome confinement free energy for the conical cavity is lower than that for the cylindrical cavity. As presented in the table, the difference is rather substantial. In part, this lowering of the confinement free energy is due to the lower overall density of segments in the cone, the volume of which is larger than a cylinder with the same surface area. { We note here that while for simplicity we employed linear chains for all the calculations in this paper, we also studied the impact of branching (following the techniques used in Ref.~\cite{Erdemci2014}) on the final results. Including the secondary structures of RNA makes the chain more compact and lower the free energy values for both cylindrical and conical shells, but do not change the conclusion of the paper.} 

\section{Gaussian chain}
\begin{figure}
\begin{center}
\includegraphics[height=8cm]{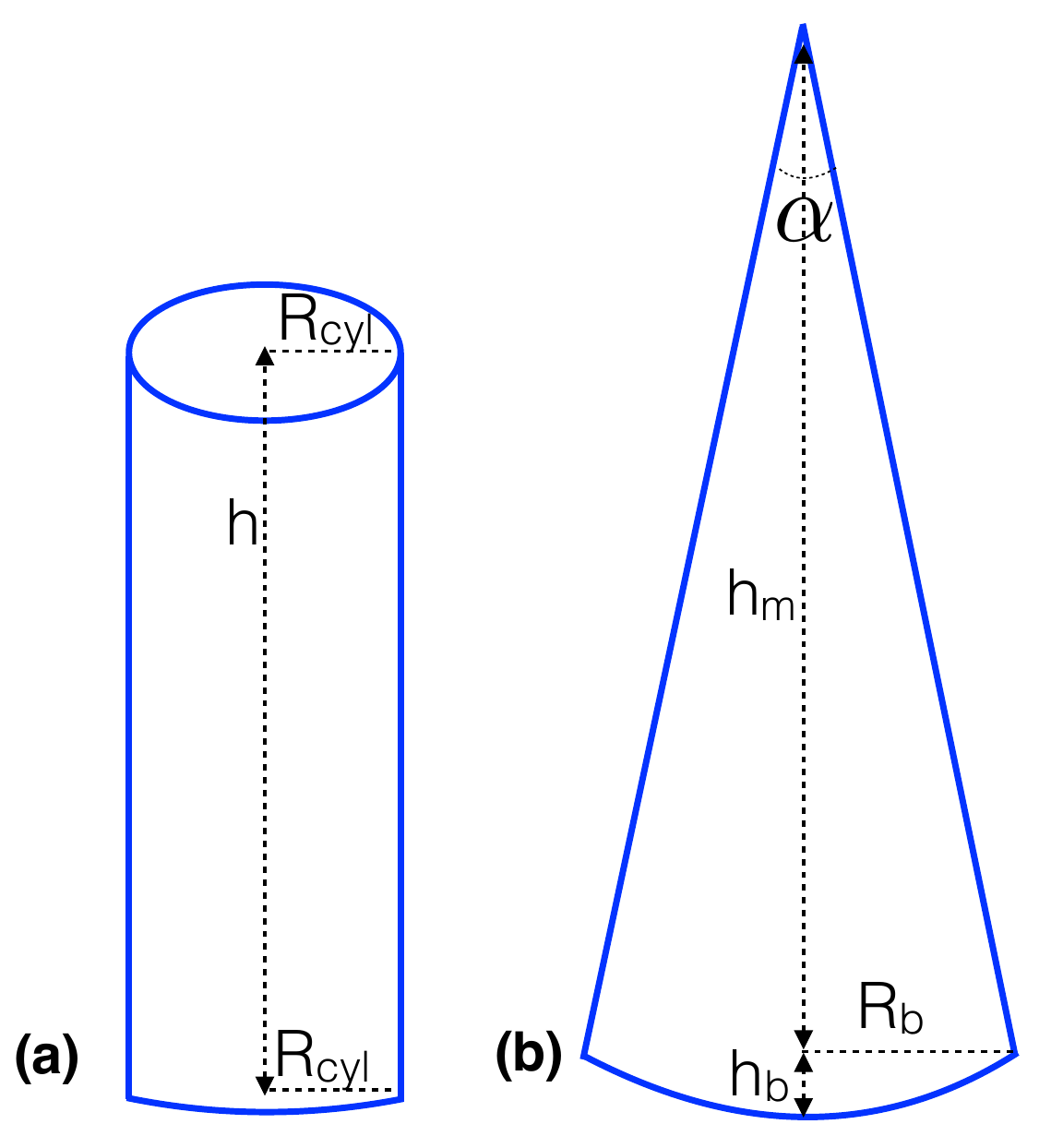}
\caption{Geometries used for the analytical calculations involving the confinement of a gaussian chain into a cylinder (a) and a cone with a sharp tip and rounded bottom (b). $R_{cyl}$ is the radius of the cylinder and $R_b$ is the bottom radius of the cone with opening angle $\alpha$. The perpendicular height of the cone is $h=h_m+h_b$ whereas the bottom cap has a height of $h_b$. The perpendicular height of the cylinder is $h$.}
\label{analytical}
\end{center}
\end{figure}

We now compare our numerical findings, presented in the previous section, with some analytical results, which we were able to find in certain limits. Indeed, it is possible to solve the Euler Lagrange equation (Eq.~\ref{diffEq}) to obtain analytical predictions for the polymer free energy and the polymer concentration profile in cylindrical and conical shells, at least if we treat the polymer as a Gaussian chain by setting $\upsilon=0$ and by assuming that the polymer does not interact with the capsid other than that it acts as a confining wall. Equation \ref{diffEq} can be solved exactly for a purely cylindrical shell, so without the caps, and for a conical structure that has a sharp tip. Both model capsids are illustrated in Fig. \ref{analytical}.

The fact that the cone of Fig.~\ref{analytical} does not quite look like the truncated cone of HIV capsids is not problematic, because we found numerically in Fig.~\ref{density-dirichlet-cone}(a) that the density of the genome for $\upsilon=0$ to be concentrated in the bottom part of the shell. Hence, the exact shape of the tip should be irrelevant to our calculations, and this is confirmed below. Similarly, the exact shape of the cylindrical cap does not effect our calculations.

For a cylindrical shell we obviously use cylindrical coordinate ${\bf x} = \{\rho, \phi, z\}$. For a cylinder of height $h$ and radius $R_{cyl}$ (Fig.~\ref{analytical}(a)), the Dirichlet boundary conditions are
\begin{eqnarray}
\psi(\rho,\phi,z=0)=\psi(\rho,\phi,z=h)&=&0,\\
\psi(\rho=R_{cyl},\phi,z)&=&0.
\end{eqnarray}
Setting $6/a^2 \lambda=\mu$, the eigenfunctions and eigenvalues of Eq.~\ref{diffEq} with $v=0$ can be found analytically. Expressed in the ``quantum'' number $n$ for the z direction, $m$ for the $\phi$ direction and $q$
for the $\rho$ direction, we find the eigenvalues and eigenfunctions
\begin{equation}\label{eq:eigenv}
\mu_{n,m,q}=\frac{\pi^2}{h^2}n^2+\frac{\lambda^2_{m,q}}{R_{cyl}^2},
\end{equation}
and
\begin{equation}\label{eq:eigenf}
\psi({\bf x}) = \sum_{n,m,q} \alpha_{n,m,q}  J_m\bigg{(} \frac{\lambda_{m,q}\rho}{R_{cyl}} \bigg{)}\sin\bigg{(}\frac{\pi}{h}nz\bigg{)}e^{im\phi},
\end{equation}
respectively. Here, $\alpha_{n,m,q}$ is the normalization factor which can be found from the constraint, Eq.~\ref{totalnumber}, and $\lambda_{m,q}$ is the $q^{th}$ zero of the $m^{th}$ Bessel function of the first kind $J_m$.

\begin{figure}
\begin{center}
\includegraphics[height=10cm]{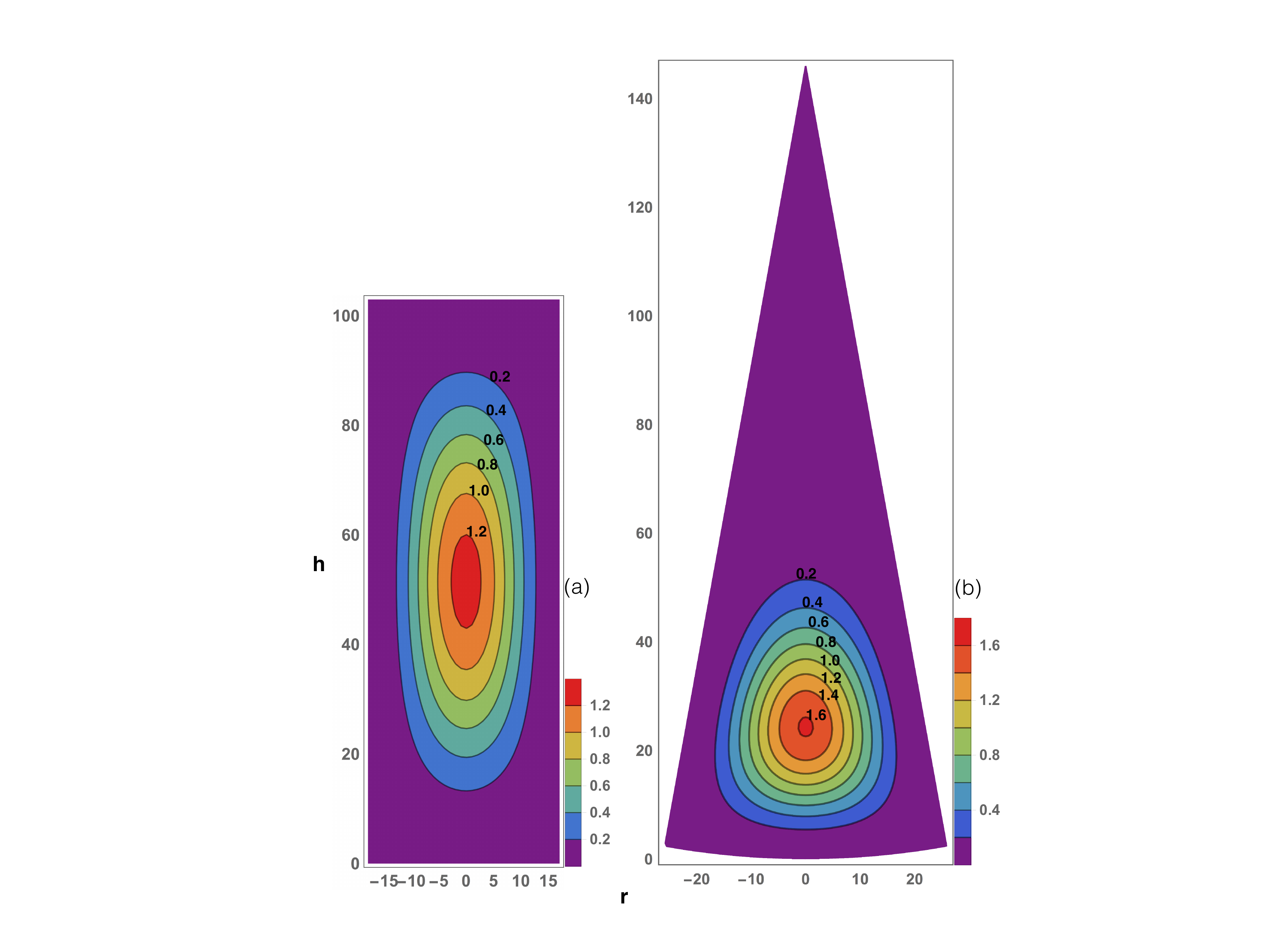}
\caption{Contour plot of the genome density profile ($\psi^2$ in units of $1/{nm^3}$) plotted from the eigenfunctions obtained from Eq.~\ref{eq:eigenf} and Eq.~\ref{eq:eigenfc} for a cylindrical (a) and conical (b) capsid. Other parameters take the values $R_b=26\;nm$, $\alpha\approx21^{\circ}$, $R_{cyl}=17.9\;nm$, $a=1\;nm$,  $h_b=2.3 \;nm$, $h=147\;nm$ for the cone and $h=103\;nm$ for the cylinder.}
\label{density-analytical}
\end{center}
\end{figure}

By inserting these expressions in Eq.~\ref{FE}, we calculate the free energy in the ground state, for long chains, so $\sqrt{N} a \gg R_{cyl}$, and we can keep only the first term in the $q$ sum. In calculating the free energy we see that all the terms besides $m=0$ integrate to zero. We also assume that the height of cylinder is very large implying $\sqrt{N} a \ll h$, and thus we can ignore the $n^2$ term in the eigenvalues. We find the free energy of a gaussian chain confined in a cylinder obeys
\begin{equation}
\beta \mathcal{F}_{cyl} \approx {\frac{N a^2}{6}\frac{\lambda^2_{01}}{R_{cyl}^2}},
\end{equation}
with $\lambda_{01} \approx 2.4$. Even if $\sqrt{N} a \gg h$, this expression holds, at least if $h \gg R_{cyl}$. 
The genome density profile inside the cylinder can be obtained from Eq.~\ref{eq:eigenf} and a plot of the density profile is presented in Fig.~\ref{density-analytical}(a).

For a polymer confined in a conical shell with a cut sphere base, depicted in Fig.~\ref{analytical}(b), it makes sense to use spherical coordinates ${\bf x} = \{r, \theta, \phi\}$. If $\alpha$ is
the opening angle of the cone and $R_b$ the radius of the spherical cap, the Dirichlet boundary conditions are
\begin{eqnarray}
\psi(r,\theta=\alpha/2,\phi)&=&0,\\
\psi(r=R_b/\sin \alpha/2,\theta, \phi)&=&0.
\end{eqnarray}
The relevant quantum numbers now become $q$ for the $r$ direction, $l$ for the
$\theta$ direction and $m$ for the $\phi$ direction. The eigenfunctions are given by
\begin{equation}\label{eq:eigenfc}
\psi({\bf x}) = \sum_{m,l,q} \alpha_{m,l,q}\; j_{\nu(l,m)}(\sqrt{\mu}_{m,l,q}r)P_{\nu(l,m)}^m(\cos \theta)e^{i m \phi}
\end{equation}
where $\alpha_{m,l,q}$ is the normalization factor which can be found from the constraint, Eq.~\ref{totalnumber}, $j_{\nu(l,m)}$ is a spherical Bessel function, $P_{\nu(l,m)}^m$ is the associated Legendre polynomial and the eigenvalues are given by
\begin{equation}
\mu_{m,l,q}=\frac{\lambda^2_{\nu(l,m)+\frac{1}{2},q}\sin^2 \alpha/2}{R_b^2}
\end{equation}
with $\nu(l,m)$ the $l^{th}$ zero of the Legendre polynomial, for which
\begin{equation}\label{Pconstraint}
P_{\nu(l,m)}^m(\cos \alpha/2 )=0
\end{equation}
holds, and where $\lambda_{\nu(l,m)+\frac{1}{2},q}$ is the $q^{th}$ zero of the Bessel function of the first kind $J_{\nu(l,m)+\frac{1}{2}}$. Plots of a spherical Bessel function with $m=0$ also indicate that the concentration of genome is maximum at the axis of the cone, consistent with numerical results for $\upsilon=0$.  Figure~\ref{density-analytical}(b) shows the location of the chain inside the conical shell, which is consistent with our numerical results (see Fig.~\ref{density-dirichlet-cone}(a)). 

To obtain the free energy of the gaussian chain confined into a conical shell, we insert Eq.~\ref{eq:eigenfc} into Eq.~\ref{FE}. In performing the integral, only the m = 0 term survives. Since the theory applies in the ground-state limit, which demands that $ \sqrt{N} a \gg R_b / \sin \alpha/2 $, we consider only the $l=1$ and $q=1$ state. We find the free energy of a gaussian chain confined in a conical shell to obey
\begin{equation}
\beta \mathcal{F}_{cone} \approx {\frac{N a^2}{6}\frac{\lambda_{\nu(1,0)+\frac{1}{2},1} \sin^2 \alpha/2}{R_b^2}},
\end{equation}
with $\nu(1,0)$ calculated from Eq.~\ref{Pconstraint} for a given $\alpha$ to obtain the value of the Bessel function zero.

In order to estimate the difference between the free energies of a gaussian chain trapped in a cylinder with respect to that trapped in a cone, we set all the relevant parameters to the values used in the numerical calculations of previous sections, {\it i.e.}, the total length of the genome $N=18000$, cone angle $\alpha\approx21^\circ$ and the radius of the base of the cone $R_b=26 \; nm$, typical for HIV particles \cite{Ganser1999,Welker2000,Briggs2003,Benjamin2005}. From this we obtain $h=147\; nm$ and the height of the spherical cap $h_b=2.3 \;nm$. We note that even though we set the height of the truncated cone in the previous sections to $h=103$, we expect that for the Gaussian chain the precise form and height of the shell are not hugely important as the genome is condensed at the bottom of the shell.  For the cylindrical capsid we set radius of the capsid at $R_{cyl}=17.9 \; nm$ and its height at $h=103\; nm$ as we did in our numerical calculations.

Based on the analytical calculation, we find that the free energy of a chain confined in the cylindrical capsid is equal to $54 \;k_B T$ and that in the conical capsid is equal to $47 \;k_B T$. These values are very close to the confinement free energy for the cylinder of $57 \;k_BT$ and for the cone of $45 \;k_BT$ that we obtained numerically.  Once again, this could explain the predominance of conical shells compared to cylindrical ones as observed in {\it in vivo} experiments.

 \section{Capsid-genome interaction}

As noted in the introduction, using the recombinant CA-NC fusion proteins and HIV-1 RNA as well as other types of RNAs, Ganser {\it et al.} obtained a mixture of cones and cylinders in their {\it in vitro} self-assembly studies \cite{Ganser1999}. We now include attractive interaction between NC proteins in the shell and RNA in the calculations. The free energy becomes \cite{Hone1988,Paul2005,Wagner2015}
\begin{equation} \label{FE2}
\beta \Delta F_{int} =-\gamma \beta a^3 \int dS \psi^2 + \beta \Delta F,
\end{equation}

\begin{figure}
\begin{center}
\includegraphics[height=8cm]{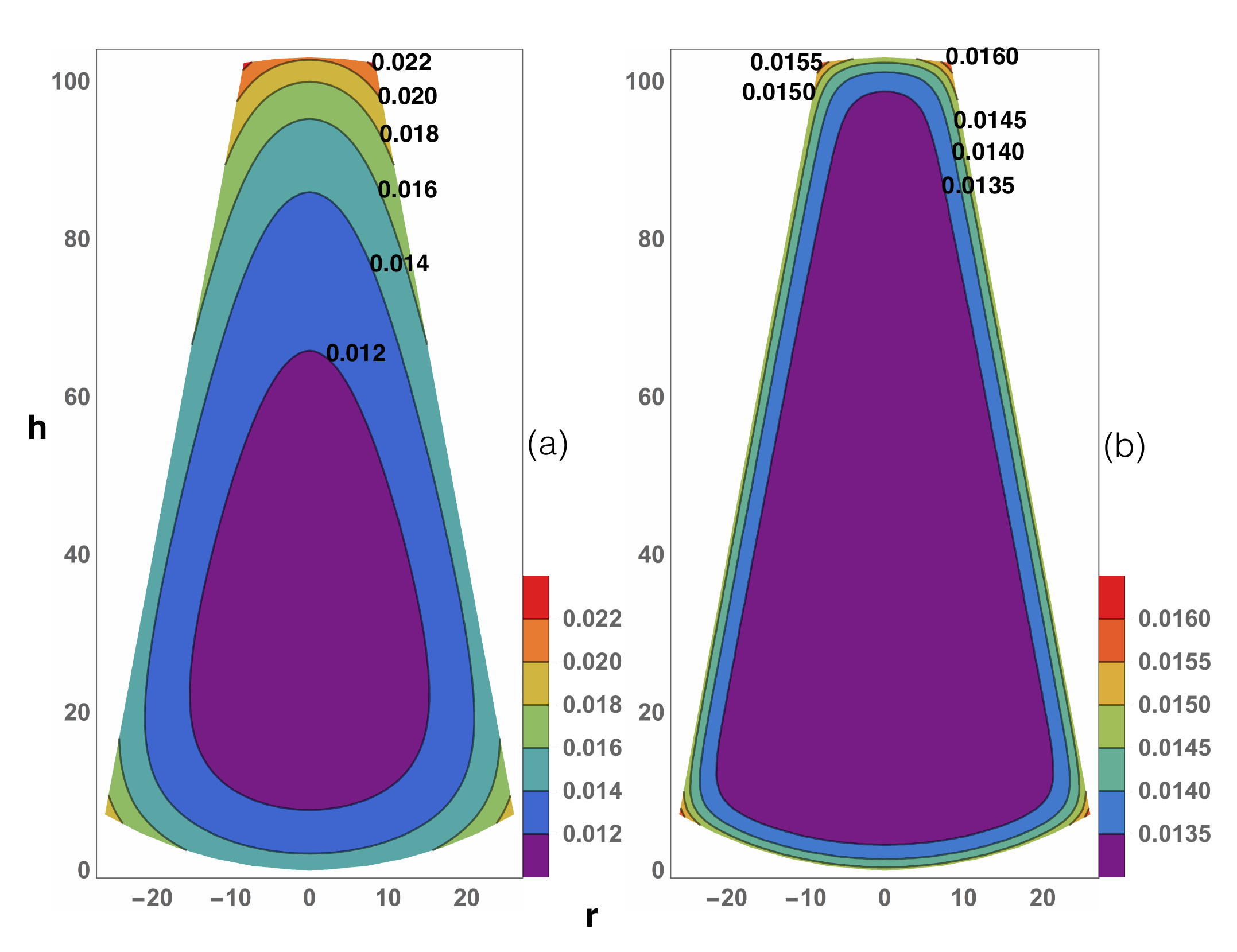}
\caption{ Contour plot of the genome density profile ($\psi^2$ in units of $1/{nm^3}$) confined into a conical capsid with an attractive inner wall for excluded volume $\nu=0.1$ (a) and $\nu=1.0 \;nm^3$ (b). The total number of monomers is $N=1400$. Other parameters are $h=103\;nm$, $R_b=26\;nm$, $\alpha=21^{\circ}$, $a=1\;nm$ and $\kappa=10^{-4} nm^{-1}$. The polymer profile is denser at the corners of the capsids, having the maximum density at the tip of the cone. }
\label{density-robin-cone}
\end{center}
\end{figure}

\begin{figure}
\begin{center}
\includegraphics[height=8cm]{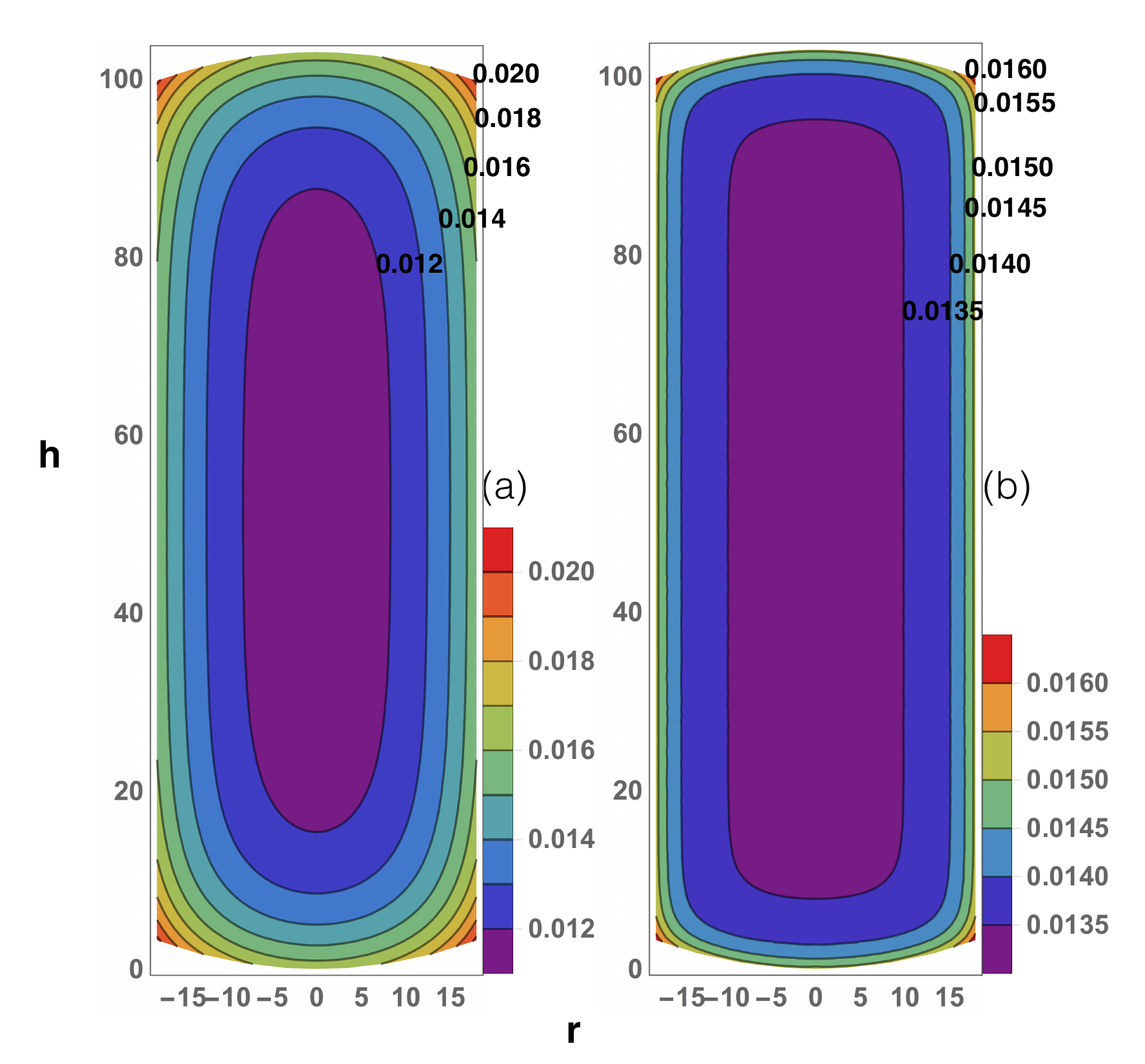}
\caption{ Contour plot of the genome density profile ($\psi^2$ in units of $1/{nm^3}$) confined into a cylindrical capsid with an attractive inner wall for excluded volume $\nu=0.1$ (a) and $\nu=1.0 \;nm^3$ (b). The total number of monomers is $N=1400$. Other parameters are $h=103\;nm$, $R_b=26\;nm$, $\alpha=21^{\circ}$, $R_{cyl}=17.9\;nm$, $a=1\;nm$ and $\kappa=10^{-4} nm^{-1}$. The polymer profile is denser at the corners of the capsids, having the maximum density at the corners of the cylinder. }
\label{density-robin-cyl}
\end{center}
\end{figure}

\noindent where $\gamma$ is the interaction energy between the genome and the inner surface of the CA-NC complex per unit area and $\Delta F$ is the polymer confinement free energy, see Eq.~\ref{FE}.
Minimizing Eq.~\ref{FE2} with respect to the field $\psi$, subject to the constraint given in Eq.~\ref{totalnumber}, produces the same Euler-Lagrange differential equation as given in Eq.~\ref{diffEq}, but subject to the following boundary conditions obtained from the minimization of Eq.~\ref{FE2} with respect to the the field $\psi$ on the surface,
\begin{equation}\label{diffEq2}
\big{(} \hat{n}\cdot \nabla \psi = \kappa \psi \big{)} \bigg{|}_{S}
\end{equation}
with $\kappa^{-1}=1/6 a \beta \gamma$ a length representing the strength of interaction of the capsid wall with the monomers. If the length $\kappa^{-1}$ is larger than the Edwards correlation length ($\zeta_{E}=a/\sqrt{3 \nu \phi_0}$) with $\phi_0=N/V$ the monomer number density, the excluded volume interaction between the monomers overcomes the monomer attraction to the wall and we are in the weak adsorption limit. Conversely, in the strong adsorption limit, $\kappa^{-1}$ is smaller than the correlation length $\zeta_{E}$ \cite{Hone1988}.

For the weak adsorption regime, corresponding in practice to a high salt concentration mimicking the experiments of Ganser {\it et al.} \cite{Ganser1999}, $\kappa^{-1}$ becomes larger than the Edwards correlation length. Figures \ref{density-robin-cone} and \ref{density-robin-cyl} show the genome density profiles in the conical and cylindrical geometries, respectively. For all plots the total number of monomers is fixed at $N=1400$, which corresponds to the HIV-1 RNA template used in the experiments of Ganser {\it et al.} \cite{Ganser1999}. The figures correspond to the weak adsorption regime with $\kappa^{-1} \gg \zeta_{E}$.

As illustrated in the contour plots for the polymer density profile in Figs.~\ref{density-robin-cone} and \ref{density-robin-cyl}, the genome covers the wall completely, and the density is higher near the rims of the cones and cylinders. This is not surprising, of course. Unlike the case studied in Sect. Model, the free energy of the genome confined in a cylindrical core, $\Delta F_{cyl}=-0.79\; k_B T$, is more negative than that of the conical core, $\Delta F_{cone}= -0.77 \; k_B T$, albeit by only a rather small amount for the excluded volume $\nu=0.1$. Since in the experiments of Ganser {\it et al.}, NC proteins are in the capsid wall and do not condense the genome, we expect a higher excluded volume interaction. For an excluded volume $\nu=1.0$, the difference between the free energies increase, {\it e.g.} $\Delta F_{cyl}=-6.62\; k_B T$ and $\Delta F_{cone}= -6.43 \; k_B T$. This would suggest, if the binding energies of the proteins are the same in both types of capsid, and kinetic effects are unimportant, then a slightly larger number of cylinders forms compared to cones, based on our calculations. Interestingly, the ratio of cylinders to cones formed in the \textit{in vitro} experiments of Ganser {\it et al.} was $3/2$, whereas the ratio of cylinders to cones found in {\it in vivo} studies is very small.

\section{Conclusions}
As noted in the introduction, CA proteins assemble spontaneously {\it in vitro} to form tubular arrays or conical structures with a geometry similar to that of the mature HIV shells, even in the absence of genome \cite{Ganser1999,Meng2012,Zhao2013}. However, Ganser \textit{et al.} showed that recombinant CA-NC proteins can assemble {\it in vitro} around any RNA to form a mixture of conical and cylindrical capsids but most structures are tubular in contrast to {\it in vivo} structures in which most capsids form conical structures. More recently, using Electron Cryotomography, Woodward {\it et al.} monitored the maturation intermediates of HIV particles  {\it in vivo} and found that most cylindrical capsids do not encapsulate RNA \cite{Woodward2015}. In fact, they found a condensed form of genome sitting next to but outside the cylindrical shells. These experiments suggest that the presence of encapsulated genome could promote the formation of conical capsids.

Several theoretical and numerical studies investigated the formation of conical structures in the absence of genome and of membrane. It seems that in the absence of genome, the formation of a conical shape could be the result of irreversible steps in the growth of an elastic sheet, and connected to the dynamics of formation of pentamers during the growth process \cite{Levandovsky2009,Wagner2015b,Jensen2013,Hagan2012}. However, the focus of this paper is solely on the impact of genome on the assembly of conical and cylindrical capsids.

Using a simple mean-field theory, we explicitly calculated the encapsulation free energy assembly of both cylindrical and conical structures. It is well-known that the interaction between the positively charged NC domain and negatively charged RNA is responsible for the encapsulation of genome in the immature HIV virus and that the CA lattice of the mature HIV does not interact with RNA.  The fact that RNA stays inside the capsid despite the absence of interaction with the CA proteins is not yet well-understood. It has been suggested that while the viral protease cleaves the CA-NC link during maturation, incomplete cleaved links could be the reason why the RNA remains encapsulated in the mature capsid \cite{Briggs2012,Woodward2015}.

It is important to note that while the interaction between CA proteins is the driving force for Gag assembly in both the mature and immature hexagonal lattices, the free energy associated with the CA-CA interaction is thought to be weak, and the free energy of the CA-CA interaction in conical and cylindrical capsids has not yet been determined experimentally. To this end, in this paper, we only focus on the contribution of genome confinement free energy, assuming that the free energies due to CA-CA interaction between cylindrical and conical capsids are not considerable as they both appear in the {\it in vitro} studies in which no genome was present.

One of the important results of this paper is that while the free energy due to confinement of genome is lower for conical capsid than a cylindrical one in {\it in vivo} experiments, where interactions between the genome and the capsid are believed to be negligible, the opposite is true if there is interaction with the wall. We emphasize that to obtain these results we employed the parameters associated with the height, radius and cone angle given in Ref.~\cite{Welker2000}. Since these parameters are not exactly the same as reported by different groups, we checked the robustness of our results using the range of parameters obtained in different experiments. {We find that our results do not depend on the exact parameters reported by one group and are robust \cite{Ganser1999,Welker2000,Briggs2003,Benjamin2005}. Furthermore, the HIV conical capsids occasionally have larger apex angles than those studied above. To this end, we compare the free energy of the encapsidated genome by a cylinder vs.~a cone with a larger apex angle than $21^{\circ}$ while keeping the area of both structures the same. Our findings turn out not to qualitatively depend on the apex angle and our conclusions remain the same.}

Understanding the factors that contribute to the formation of conical capsids can play an important role in the development of anti-viral drugs and nano-containers for gene therapy.

\label{discussion}

\section*{Acknowledgments}
The authors thank William Gelbart for many useful and productive discussions over many years on the physics of virus assembly, including the work presented in this paper. This work was supported by the National Science Foundation through Grant No. DMR-13-10687 (R.Z.).

\bibliography{conical}
\end{document}